\documentclass{aastex6}

\newcommand{\oql}{O$^{7+}$/O$^{6+}$}
\newcommand{\feo}{$N_{Fe}/N_O$}
\newcommand{\velunit}{km~s$^{-1}$}

\begin{document}


\title{Charge states and FIP bias of the solar wind from coronal holes,
    active regions, and quiet Sun}


\author{Hui Fu\altaffilmark{1}, Maria S. Madjarska\altaffilmark{2,1}, LiDong Xia\altaffilmark{1}, Bo Li\altaffilmark{1}, ZhengHua Huang\altaffilmark{1}, Zhipeng Wangguan\altaffilmark{1}}


\altaffiltext{1}{Shandong Provincial Key Laboratory of Optical
Astronomy and Solar-Terrestrial Environment, Institute of Space Sciences, Shandong University, Weihai 264209, Shandong,China; xld@sdu.edu.cn}
\altaffiltext{2}{Max Planck Institute for Solar System Research, Justus-von-Liebig-Weg 3, 37077, G\"ottingen, Germany}


\begin{abstract}
Connecting in-situ measured solar-wind plasma properties with typical regions
    on the Sun can provide an effective constraint
    and test to various solar wind models.
We examine the statistical characteristics of the solar wind with an origin
    in {  different types of source regions}.
We find that the speed distribution of coronal hole (CH) wind is bimodal
    with the slow wind peaking at $\sim$400~\velunit\ and a fast at $\sim$600~\velunit.
    An anti-correlation between the solar wind speeds and the \oql\ ion ratio
    remains valid in all three types of solar wind as well during  the three studied solar cycle activity phases, i.e. solar maximum, decline and minimum.
The \feo\ range and its average values all
    decrease with the increasing solar wind speed in different types of solar wind.
The \feo\ range (0.06--0.40, FIP bias range 1--7) for AR wind is wider than for CH wind (0.06--0.20, FIP bias range 1--3) while the minimum value of \feo\ ($\sim$~0.06) does not change with
    the variation of speed,
    and it is similar for all source regions.
The two-peak distribution of CH wind and
    the anti-correlation between the speed and \oql\
    in all three types of solar wind can be explained qualitatively
    by both the wave-turbulence-driven (WTD) and
    reconnection-loop-opening (RLO) models,
    whereas the distribution features of \feo\ in different source regions
    of solar wind can be explained more reasonably by the RLO models.

\end{abstract}

\keywords{Solar wind: sources -- Solar wind: charge states -- Solar wind: FIP bias -- Solar wind: heating and acceleration}



\section{Introduction} \label{sec:intro}

It is  common knowledge that the in-situ solar wind has two basic components:
     a steady fast ($\sim$800~\velunit)  and a variable slow ($\sim$400~\velunit) component
    \citep[e.g.][and the references therein]{2006SSRv..124...51S}.
While it is widely accepted that the fast solar wind (FSW) originates
    in coronal holes
    \citep{1973SoPh...29..505K,1977RvGSP..15..257Z,1999SSRv...89...21G},
    the source regions of the slow solar wind (SSW) are still poorly understood.
One of the sources of the SSW has been linked to sources at the edges
    of active regions
    \citep[e.g.][etc.]{1999JGR...10416993K, 2007Sci...318.1585S, 2014SoPh..289.3799C}
    and it is also believed that the SSW originates in the quiet Sun
    \citep[e.g.][]{2000JGR...10512667W, 2005JGRA..110.7109F, 2015SoPh..290.1399F}.

Intuitively, identification of wind sources can be done by tracing
    wind parcels back to the Sun.
By applying a potential-field-source-surface (PFSS) model,
    \citet{2002JGRA..107.1154L}
    mapped a low latitude solar wind back to the photosphere for nearly
    three solar activity cycles. They showed, for instance, that polar coronal holes contribute to the solar wind only over about the half solar cycle, while for the rest of the time the low-latitude solar wind  originates from ``isolated low-latitude
and midlatitude coronal holes or polar coronal hole
extensions that have a flow character distinct from that
of the large polar hole flows''. Using a standard two-step mapping procedure,
    \citet{2002JGRA..107.1488N}
    traced  solar wind parcels back to the solar surface
    for four Carrington rotations during a solar maximum phase.
The solar wind was divided into two categories: coronal hole and
    active region wind, and their statistical parameters were analyzed
    separately.
The authors reported that the \oql\ ion ratio is lower for the coronal-hole wind
    in comparison to the active-region wind.
\citet{2015SoPh..290.1399F}
    traced the solar wind back to its sources and classified
    the solar wind by the type of the source region,
    i.e. active region (AR), quiet Sun (QS) and coronal hole (CH) wind.
They found that the fractions occupied by each type of solar wind
     change with the solar cycle activity and  established that the quiet Sun regions are an important source of the solar wind during the solar minimum phase.

Alternatively, wind sources can be determined by examining in-situ charge
    states and elemental abundances.
For the former, the charge states of species such as oxygen and carbon are regarded
    as a telltale signature of the solar wind sources.
For example, the {density ratio of n(O$^{7+}$) to n(O$^{6+}$) (i.e. ionic charge state ratio, hereafter  \oql)} does not
    vary with the distance
    beyond several solar radii above the solar surface, and, therefore,
    it reflects the electron temperature in the coronal sources
    \citep{1983ApJ...275..354O,1986SoPh..103..347B}.
As the temperatures in different source regions are different,
    therefore, the source regions can be identified by the charge states
    detected in situ
    \citep{2009GeoRL..3614104Z, 2012ApJ...744..100L}.
The first ionization potential (FIP) effect describes the element anomalies in the
    upper solar atmosphere and the solar wind (especially in the SSW),
    i.e. the abundance increase of elements with a FIP of less than 10 eV (e.g.,  Mg, Si, and Fe) to those with a higher FIP (e.g., O, Ne, and He).
The in-situ measured FIP bias is usually represented by \feo\,
    and it can be expressed as
    \begin{equation}
    FIP~bias=\frac{(N_{Fe}/N_O)_{solar~wind}}{(N_{Fe}/N_O)_{photosphere}},
    \end{equation}
    where $N_{Fe}/N_O$ is the abundance ratio of iron (Fe) and
    oxygen (O).
In the slow wind the FIP bias is  $\sim$3,
    while in the fast streams it is found to be smaller
    but still above 1
    \citep{2000JGR...10527217V}.
As the FIP bias in coronal holes, the quiet Sun, and active regions
    has significant differences, the solar wind detected in situ
    can be linked to those source regions
    \citep[e.g.,][and the references thererin]{2005JGRA..110.7109F, 2015LRSP...12....2L}.
In the present study, we only present the in-situ measurements of $N_{Fe}/N_O$
    that can easily be translated to a FIP bias by considering $N_{Fe}/N_O$ in the photosphere to be a constant  at $\sim$0.06 \citep{2009ARA&A..47..481A}.

Two theoretical frameworks, the wave-turbulence-driven (WTD) models and
    the reconnection-loop-opening (RLO) models, have been proposed
    to account for the observational results.
In the WTD models, the magnetic funnels are jostled by the
    photosphere convection,
    and waves are produced that propagate into the upper atmosphere.
These waves can dissipate to heat and
    accelerate the nascent solar wind
    \citep{1986JGR....91.4111H, 1991ApJ...372L..45W, 2007ApJS..171..520C, 2009EM&P..104..121V}.
In the RLO models the magnetic-field line of loops reconnect with open field lines and
    during this process mass and energy are released
    \citep{1999JGR...10419765F, 2003JGRA..108.1157F, 2003ApJ...599.1395S, 2004ApJ...612.1171W, 2006JGRA..111.9115F}.
For more details on these two classes of models please see the review by
    \citet{2009LRSP....6....3C}.
There are two important differences between the two models.
First, in the WTD models the plasma escapes directly along open magnetic-field lines,
    whereas in the RLO models the plasma is released from closed loops
    through magnetic reconnection.
Second, in the WTD models the speed of the solar wind is determined
    by the super radial expansion
    \citep{1990ApJ...355..726W}  and curvature degree   \citep{2011A&A...529A.148L}
    of the open magnetic-field lines.
In the RLO models, the speed of the solar wind depends on
    the {  temperature} of the loops that reconnect with
    the open magnetic-field lines
    with {  hotter} loops producing slow wind,
    and {  cooler} loops producing fast wind
    \citep{2003JGRA..108.1157F}.

Observations can be used to test any of the above mentioned models.
For this purpose, the following three questions need to be addressed.
First, where do the two components (a steady fast component and
    a variable slow portion
    \citep{2006SSRv..124...51S})
    of the solar wind originate from?
Second, why does the charge state anti-correlate with the solar wind speed
    \citep{1995SSRv...72...49G, 2000JGR...10527217V, 2003JGRA..108.1158G, 2003ApJ...587..818W, 2009ApJ...691..760W}?
Third, why is the FIP bias  (FIP bias value range) higher (wider)
    in the slow solar wind than in the fast wind
    \citep{1995SSRv...72...49G, 2000JGR...10527217V, 2016SSRv..201...55A}?

Traditionally, solar wind is classified by its speeds.
The speed, however, is not the only characteristic feature of
    the solar wind
    \citep{2012SSRv..172..169A, 2016SSRv..201...55A}.
The plasma properties and magnetic-field structures can be
    significantly different depending on the solar regions,
    i.e. coronal holes, quiet Sun, and active regions.
The differences in the source regions would then influence the solar wind
    streams they generate \citep{2005JGRA..110.7109F}.
Therefore, connecting in-situ measured solar-wind plasma properties with
    typical regions on the Sun can provide an effective
    constraint and test to various solar wind models
    \citep{2014ApJ...790..111L}.
In an earlier study,
    we classified the solar wind by the source region
    type (CHs, QS, and ARs)
    \citep{2015SoPh..290.1399F}.
Here, we analyze the relationship between  in-situ solar wind parameters and
    source regions in different phases of the solar cycle activity.
We aim at answering the following outstanding questions:
1) Are there any differences in the speed, \oql, and \feo\ distributions
    of the different types of solar wind, i.e. AR, QS and CH?
2) Is the anti-correlation between the solar wind speed and
    \oql\ charge state still valid for each type of solar wind?
3) What are the characteristics of the distribution in speed and \feo\ space
    for the different types of solar wind?
We also discuss our new results in the light
    of the WTD and RLO models.

The paper is organized as follows.
In Section~2 we describe the data and the analysis methods.
The statistical results are discussed in Section~3.
The summary and concluding remarks are given in Section~4.

\section{Data and Analysis} \label{sec:data}

In \citet{2015SoPh..290.1399F}, the source regions were categorized
    into three groups:
    CHs, ARs, and QS,
    and the wind streams originating from these regions were given
    the corresponding names
    CH wind, AR wind, and QS wind, respectively. We used  hourly averaged solar wind speeds
    measured by  the \textit{Solar Wind Electron, Proton, and Alpha Monitor}
    onboard the \textit{Advanced Composition Explorer}
    \citep[ACE,][]{1998SSRv...86....1S}.
The charge state \oql\ and FIP bias \feo\  (also hourly averaged)
    were recorded by the
    \textit{Solar Wind Ion Composition Spectrometer},
    \citep[ SWICS/ACE][]{1998SSRv...86..497G}.
In the present study we are only interested in the non-transient solar wind, and therefore,
    the intervals occupied by Interplanetary
    Coronal Mass Ejections (ICMEs) were excluded.
We used  the method suggested by \citet{2004JGRA..109.9104R} where charge state \oql\ exceeding
    $6.008 exp(-0.00578v)$ ($v$ is the ICME speed) were  discarded.
In this study, the threshold for FSW and SSW is chosen as 500~\velunit\ \citep{2015SoPh..290.1399F, 2016SSRv..201...55A}.

{\ {
The two-step mapping procedure
    \citep{1998JGR...10314587N, 2002JGRA..107.1488N}
    was applied  to trace the solar wind parcels back to the solar surface.
The footpoints were then placed on the EUV images observed by the
    Extreme-ultraviolet Imaging Telescope (EIT, \citeauthor{1995SoPh..162..291D},
    \citeyear{1995SoPh..162..291D})
    and the photospheric magnetograms taken by the
    Michelson Doppler imager
    (MDI, \citeauthor{1995SoPh..162..129S} \citeyear{1995SoPh..162..129S})
    onboard SoHO
    (Solar and Heliospheric Observatory,
    \citeauthor{1995SoPh..162....1D}, \citeyear{1995SoPh..162....1D}).
Here, the EIT~284~\AA\ passband was used as coronal holes are best distinguishable there.}}

{\ {
The scheme for classifying the source regions is illustrated
    in the  top panels (a--d) of Figure~\ref{fig:fig_parameter_2003}
    where the footpoint locations (red crosses) are overplotted
    on the EIT images (a1, b1, c1, d1) and the photospheric magnetograms  (a2, b2, c2, d2).
The wind with footpoints located within CHs is classified as ``CH wind".
A quantitative approach which follows that of \citet{2009SoPh..256...87K}
    for identifying coronal hole boundaries is implemented.
In this approach, a rectangular box which includes apparently dark area
    and its brighter surrounding area is chosen.
There would be a multipeak distribution for its intensity histogram
    (see Figure~3 in \citet{2009SoPh..256...87K} and Figure~2 in \citet{2015SoPh..290.1399F}).
The minimum between the first two peaks was defined
    as the threshold for the CH boundary (see the green contours in Figure~\ref{fig:fig_parameter_2003}, a1--d1).
This scheme can define CH boundaries more objectively and it is not influenced
    by the emission variation of the corona with the solar activity.
The definition of the AR wind  relies on the magnetic field strength
at photospheric level and corresponds to magnetically concentrated areas (MCAs). MCAs conform an area  defined by the value of contour levels
  that are 1.5--4 times the mean of the radial component
    of the photospheric magnetic field.
We found that the morphology of MCA is not sensitive to contour levels
    if it is in the above mentioned range.
This means that the MCAs have a strong spatial gradient of the radial  magnetic-field component.
The defined MCAs encompass all active regions numbered by NOAA as given
    by the solarmonitor \footnote{http://solarmonitor.org}.
However, not all MCAs correspond to an AR numbered by NOAA.
As CH boundaries are defined quantitatively, we need only to consider
    the regions outside CHs when we identify AR and QS regions.
An AR wind is defined when its footpoint is located inside an MCA that
    is a numbered NOAA AR.
The QS wind is defined when the footpoints are located outside any MCA and CH.
The  regions  for which a footpoint is located  in MCA that is not numbered by
    NOAA are named as ``Undefined" in order to keep the selection of the three groups solar wind ``pure''.
The fractions of the undefined group range from $\sim$~5\% to $\sim$~20\%
    for the years 2000 to 2008.
More details on the background work can be found in
    \citet{2015SoPh..290.1399F}.
}}

{\ {
In the present study, the temporal resolution of the data used
    for tracing the solar
    wind back to the solar surface is enhanced to 12 hours.
{
The data for which the polarities are inconsistent at the two ends are removed
    as done in \citet{2002JGRA..107.1488N} and \citet{2015SoPh..290.1399F}.}
The statistical results for the solar wind parameters are almost the same as
  those of \citet{2015SoPh..290.1399F}
    in which the temporal resolution is 1-day.
More detailed analysis shows that
    the footpoints stay in a particular region (CH, AR or QS)
    for several days.
One example is shown in Figure~1 (a1), where a footpoint is located in
    the same big equatorial hole for almost 7 days.
This means that a higher temporal resolution can only influence
    the classification when a footpoint lies near the edge of a certain region.
    The analysed data cover the time period from 2000 to 2008 which is further
    divided into a solar maximum (2000--2001), decline (2002--2006),
    and minimum phases (2007--2008)
    based on the monthly sunspot number.

\section{Results and discussion} \label{sec:results}
\subsection{Parameter distributions}
\label{subsec:paradistribution}
To demonstrate the linkage of in-situ measured solar wind speeds,
    \feo\ and \oql\ to particular solar regions,
    we describe in detail a randomly chosen example of a period of time with  typical CH, AR, and QS wind.
{\ {Figure~\ref{fig:fig_parameter_2003} provides an illustration
    of the classification scheme
    of the solar wind (a--d) and}
    the solar wind parameters
    for the time period from day 313 to day 361 of 2003 (e--g).
The footpoint of solar wind parcel detected by ACE on day 316
    is shown in Figure~\ref{fig:fig_parameter_2003}, a1 and a2, day 336 corresponds to b1 and b2,
    day {  341} to c1 and c2, day 346 to d1 and d2.
The solar wind was classified as CH, AR, QS and CH wind, respectively.
}
The period of time from day 315 to day 322 and 345 to 349 show
    two fast solar wind streams with an average speed of
{ \ { $\sim$700~\velunit\ and $\sim$800~\velunit.}}
These two streams are separated {  approximately 27 days}
    during which the streams
    from two active regions, an {\ equatorial} coronal hole
    and two quiet-Sun periods are identified.
The two fastest streams are associated with a low charge state
    and a \feo\ ratio of
   {\  0.03$\pm$0.015} and {\  0.1$\pm$0.01}, respectively.
{\
As shown in Figure~\ref{fig:fig_parameter_2003}, d1, the fast stream
    (from day 345 to day 349 ) originates from a big equatorial coronal hole.
Thus, {  the start (days 342.0-343.5) and end (days 350.5 to 352.0)} periods
    are considered as coronal-hole-boundary origin regions.
They have parameters characteristic for slow solar wind, i.e. {the
    charge state \oql\ is 0.14$\pm$0.07 and 0.13$\pm$0.03, and
    \feo\ -- 0.20$\pm$0.04 and 0.14$\pm$0.01}, respectively.
}
{In contrast to the two fastest CH streams}, the AR wind has a lower speed of $\sim$500~\velunit\ and
    $\sim$400~\velunit, and higher \oql\ and \feo\
    {(0.20$\pm$0.05 and 0.29$\pm$0.05, and 0.15$\pm$0.01 and 0.14$\pm$0.02,
    respectively). }
The QS wind has parameters comparable to the AR speeds ($\sim$450~\velunit),
    {   \oql\ (0.19$\pm$0.04 and 0.18$\pm$0.05) and \feo\ ratios (0.10$\pm$0.01 and 0.12$\pm$0.02).
The speed, \oql\ and \feo\ ratios for the wind that originates from the
    equatorial CH (day 327.5-332.5) are
    524$\pm$80~\velunit, 0.10$\pm$0.04 and 0.12$\pm$0.01.}

\begin{figure*}[ht!]
\figurenum{1}
\centerline{\includegraphics[width=0.95\textwidth,clip=]
{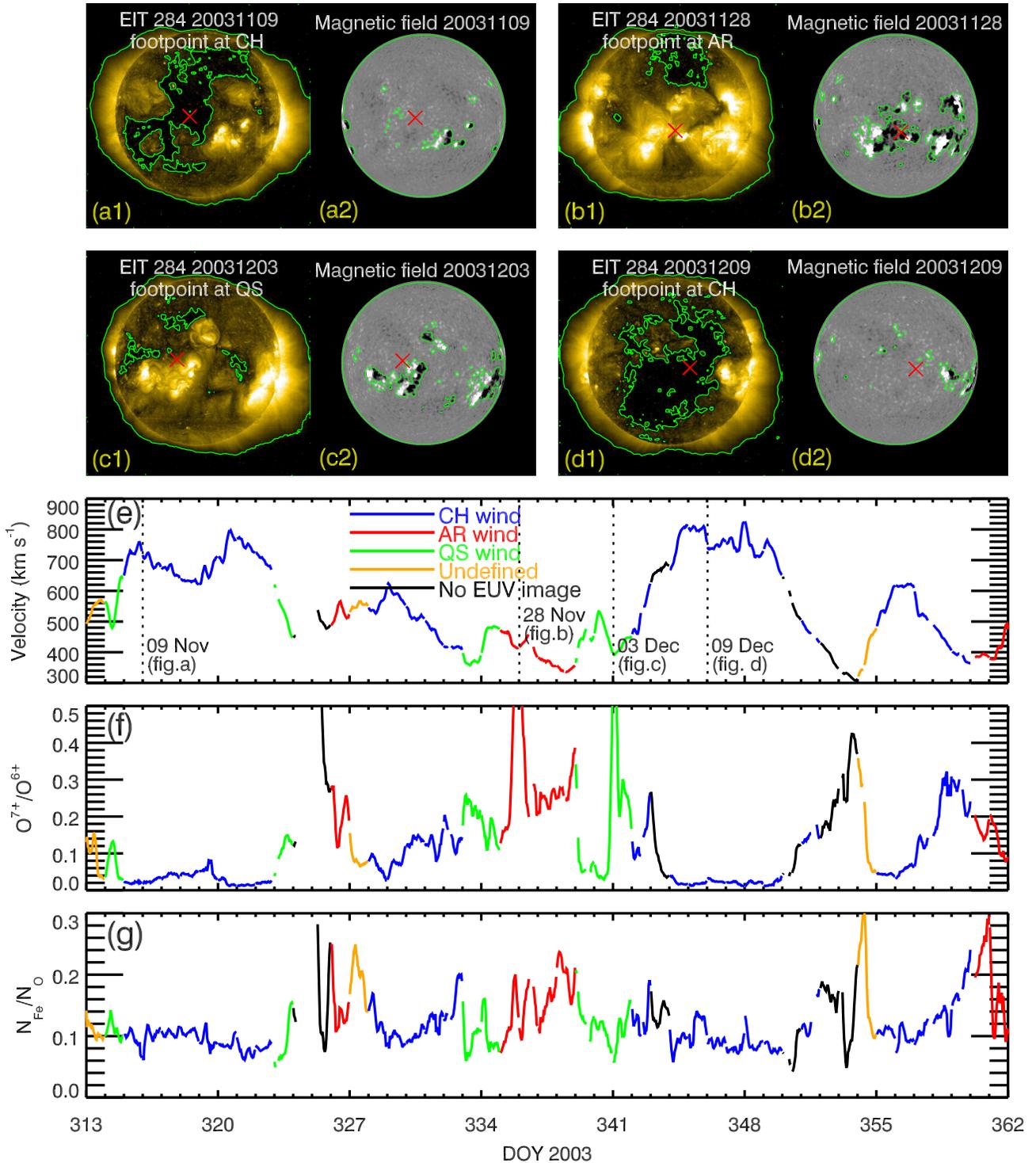}}
\caption{
\ {
Top panel (a--d) illustrates the classification scheme of the solar wind.
Figures~a1, b1, c1 and d1 present the EIT 284 \AA\ images,
    while a2, b2, c2, d2 give the corresponding photospheric magnetograms,
with green contours outlining the CHs and MCA  boundaries, respectively.
The footpoints of the solar wind are denoted by {  red} crosses.
The solar wind detected by ACE is classified as CH wind (a1, a2 and d1, d2),
    AR wind (b1, b2) and QS wind (c1, c2).
Bottom panels: Solar wind parameters for days 313--361 of 2003.
Panel (e)--(g) show the speeds, \oql, and \feo\.
The wind streams from CHs, ARs, and the QS are represented by
    blue, red, and green lines, respectively.
The orange lines denote the undefined wind type,
    and the black lines represent the days during which
    there is no EUV image taken by EIT.
The vertical lines represent the solar wind flows whose footpoint is shown
    in figure (a)--(d), respectively.
}
\label{fig:fig_parameter_2003}}
\end{figure*}

\begin{table*}
\center
\caption{The contribution of different source regions to FSW and SSW
    \ {  determined from the ACE measurements}
    during a given solar cycle phase.}
\label{tbl_1}
\begin{tabular}{c|ccc|ccc}     
  \hline                   
  \hline
   &        &  FSW   &        &        &  SSW   &        \\
  \hline
   & CH     &   AR   &   QS   &   CH   &   AR   &   QS   \\
  \hline
ALL & 39.3\% & 25.2\% & 35.5\% & 13.4\% & 42.9\% & 43.7\% \\
MAX & 34.1\% & 40.3\% & 25.6\% & 12.1\% & 58.8\% & 29.1\% \\
DEC & 48.2\% & 24.0\% & 27.8\% & 16.1\% & 41.5\% & 42.3\% \\
MIN & 23.2\% & 12.4\% & 64.3\% & 11.2\% & 15.9\% & 72.9\% \\

  \hline
\end{tabular}
\end{table*}

\begin{table*}
\center
\caption{The proportion of FSW and SSW
    \ { determined from the  ACE measurements}
    for a given source region during
    different solar cycle phases.}
\label{tbl_2}
\begin{tabular}{c|cc|cc|cc|cc}     
  \hline                   
  \hline
   &\multicolumn{2}{c|}{ALL}& \multicolumn{2}{c|}{MAX}&\multicolumn{2}{c|}{DEC}&\multicolumn{2}{c}{MIN}\\
  \hline
   & FSW    &   SSW  &  FSW   &  SSW   &  FSW   &  SSW   &  FSW   &  SSW   \\
  \hline
CHs & 59.3\% & 40.7\% & 39.5\% & 60.5\% & 64.3\% & 35.7\% & 57.8\% & 42.2\% \\
ARs & 22.6\% & 77.4\% & 13.7\% & 86.3\% & 25.8\% & 74.2\% & 34.0\% & 66.0\% \\
QS & 28.8\% & 71.2\% & 17.0\% & 83.0\% & 28.4\% & 71.6\% & 36.8\% & 63.2\% \\
  \hline
\end{tabular}
\end{table*}

As the magnetic field configurations and plasma properties
   are significantly different for the different types of source regions,
   we investigate here whether there also exist differences
   between the three {  sources of solar wind}.
In Figure~\ref{fig:fig_parameter_histrogm},
    we show the {\  normalized (to the maximum for each wind type)} distributions
    of the solar wind speed and the \oql\ and \feo\ ratios
    for the solar wind as a whole (in order to compare with earlier studies) and the different source regions, i.e. AR, QS and CH.
As these parameters are expected to change with the solar activity
    \citep{2013ApJ...768...94L, 2015SoPh..290.1399F},
    the results are shown for three different phases of solar cycle 23.
From Figure~\ref{fig:fig_parameter_histrogm} (a1) we note that
    as expected the average speed of the CH wind is higher than the AR and QS wind.
Further, we estimated the contribution of each of the source regions to the fast
    and the slow solar wind (Table~\ref{tbl_1}).
If the whole cycle is considered {  as} one,
    the CHs (39.3\%) have just a $\sim$4\% higher contribution to the FSW than the QS (35.5\%), with ARs having the smallest input of 25.2\%.
For a given solar cycle phase, however, the true contribution of each type of
    solar wind becomes more evident.
During the maximum phase of the solar cycle, the ARs are the dominant FSW source
    at 40.3\% followed by the CHs at 34.1\% and the QS at 25.6\%.
At the time of the decline phase the CHs are prevailing  at {  48.2\%} and
    the rest of the FSW input comes almost equally from the QS (24.0\%) and ARs (27.8\%).
The most dominant during the minimum is the QS at 64.3\%.
With regard to the SSW, if all the whole cycle is examined,
    the {  QS (43.7\%) and ARs (42.9\%) have an almost equal contribution.}
At the maximum ARs are the main source of SSW at 58.8\%,
    while in the decline phase again the AR and QS have
    almost the same {  contribution of $\sim$42\%}.
The predominant source of the SSW during the minimum of the solar activity
    is the QS at 72.9\%.

The fractional contribution of each wind for a given source region is shown in Table~~\ref{tbl_2}.  Again if all solar cycle phases together are studied, then CHs produce $\sim$60\% FSW ($\sim$40\% SSW), ARs $\sim$77\% SSW of their total solar wind input, while $\sim$29\% of the total QS contribution goes into the FSW. More interesting is, however, how the fractions change during the different phases of the solar cycle activity. During the maximum CH SSW rises to $\sim$60\%,  while the solar  main contribution  of ARs and QS (of more than $\sim$80\%) is to the SSW. In the decline phase the CH { produces} more FSW ($\sim$64\%) than SSW ($\sim$36\%). Again the ARs and QS are predominantly contributing to the SSW at a bit more than $\sim$70\% of their total wind contribution.  During the minimum the CH FSW contribution decreases slightly to $\sim$58\% while the emission of SSW grows to $\sim$42\%. In the minimum the FSW contribution of the ARs and QS goes further up to ~34\% and ~37\%, respectively, while their SSW contribution decrease.
{\
It is important to point out that the above results only reflect the wind detected
    by ACE which lies in the ecliptic plane.
{  Also we have to note that the heliospheric structures
    (such as neutral line and heliospheric current sheet)
    which may influence the statistical results were not removed during the investigated years.
Usually, those structures are associated with the boundaries between
 different source regions of the solar wind
    \citep{2002JGRA..107.1488N}.}
}

Case and statistical studies have already shown that CHs are sources of
    both the fast and slow wind
    \citep{2002JGRA..107.1488N, 2014ApJ...787..121K}.
    \citet{2000JGR...10512667W}
    compared the flow speed derived from Doppler dimming and
    density observations by UVCS/SoHO (UltraViolet Coronagraph Spectrometer)
    and suggested that the QS regions are an additional source
    of the fast solar wind, thus questioning the
    {\  traditional belief}
    that the fast solar wind originates only from CHs.
 {\ {
There are many small regions (size of  several arcsec across) in a typical QS region
    that are similar in brightness to CH regions as observed in, for instance,  coronal spectral lines Ne~{\sc viii} (T$_{max}$ $\sim$ 6$\times$10$^5$~K)  and Mg~{\sc x}  (T$_{max}$ $\sim$ 10$^6$~K) (from SUMER), as well as EIT and TRACE coronal solar-disk images.
Thus, \citet{2005JGRA..110.7109F} speculated that those dark QS regions
    may be the source regions of the fast solar wind suggested by
    \citet{2000JGR...10512667W}.
}}
In the present study, the solar wind is classified by source regions which differs from classifications that are based on solar wind
    parameters (such as solar wind speed and charge state \oql) \citep{2009GeoRL..3614104Z, 2012ApJ...744..100L}.
Bearing the uncertainties of our solar-wind classification scheme as discussed
    in \citet{2015SoPh..290.1399F} (such as the reliability of the PFSS model and the simple ballistic treatment in the mapping procedure),
our results  demonstrate the complexity of the fast and slow
    solar wind origin.


A significant feature for the CH wind speeds are their two peak
    distributions for all three solar activity phases.
We  {\  suggest} that the fast and slow distribution peaks
    come from the CH center and the boundary regions that
    include both CH open and QS/AR close magnetic fields, respectively.
For instance, as shown in the wind-source identification example in
    Figure~\ref{fig:fig_parameter_2003},
    the solar wind streams coming from CH core regions are faster,
    while the CH boundary streams are slower.
Similar results are shown by
    \citet[][see Figure 11 in their paper]{2002JGRA..107.1488N} and
    \citet[][Figure~1 in their paper]{2014ApJ...787..121K}.
 Several studies have suggested  that structures like
    coronal bright points and plumes which are located in
    CH boundary regions may also be sources of the SSW
    \citep[e.g.,][]{2004ApJ...603L..57M, 2010A&A...516A..50S, 2012A&A...545A..67M,
    2014ApJ...794..109F, 2016arXiv160609201K}.
This can be interpreted by both the WTD and RLO models.
In the WTD models, it means that the super radial expansion and curvature
    degree of the open magnetic-field lines are smaller and lower at  the center of CHs compared to CH boundary regions.
RLO models suggest that loops that reconnect with open
    magnetic-field lines have {  lower temperature} at CH central regions than loops located in the boundary regions.
The two peak distribution of solar wind speeds may, therefore, reflect either
    the super radial expansion and the curvature degree
    of the open magnetic-field lines, or loop {  temperature}.

As shown in Figure~\ref{fig:fig_parameter_histrogm}, (b1) the CH solar-wind
    speed distribution has a stronger peak at $\sim$400~\velunit,
    and a second weaker peak at $\sim$600~\velunit\
    during the solar maximum.
During the decline and minimum phases,
    the second peak at $\sim$600~\velunit\ is stronger than the peak at lower speeds.
{\  The two-peak distribution variation of the CH wind may be related to the
    different average areas and physical properties (such as magnetic field strength)
    of CHs during different solar activity phases \citep{2009SSRv..144..383W}.
Another possible explanation is the difference of the heliospheric structure
    during the different solar cycle phases.
During solar maximum, ACE may encounter a longer period of neutral line or
    heliospheric current sheet which usually corresponds to the SSW,
thus  the slow wind peak is higher.
{  The faster peak at $\sim$600~\velunit\ is stronger
    because there are more
    equatorial CHs during the decline phase
    \citep{2009SSRv..144..383W}}.


As it can be seen from the middle and bottom rows in
    Figure~\ref{fig:fig_parameter_histrogm},
    in all three phases of the solar cycle discussed here the average values of \oql\ and \feo\ are the highest  for the AR wind,
    the lowest for the  CH wind  with the QS wind in between.
Consistent with
    \citet{2016SoPh..291.2441K},
    our study demonstrates that the distributions of the
    speed, \oql, and \feo\ ratios for the different types of solar wind
    have a large overlap, and therefore,
    it is hard to distinguish the source regions only by those wind parameters.
The other characteristics for the parameters of \oql\ and \feo\ ratio will be
    discussed in Section~\ref{subsec:spacedistribution_vel_o76} and \ref{subsec:spacedistribution_vel_feo}

\begin{figure*}[ht!]     
\figurenum{2}
\centerline{\includegraphics[width=0.95\textwidth,clip=]
{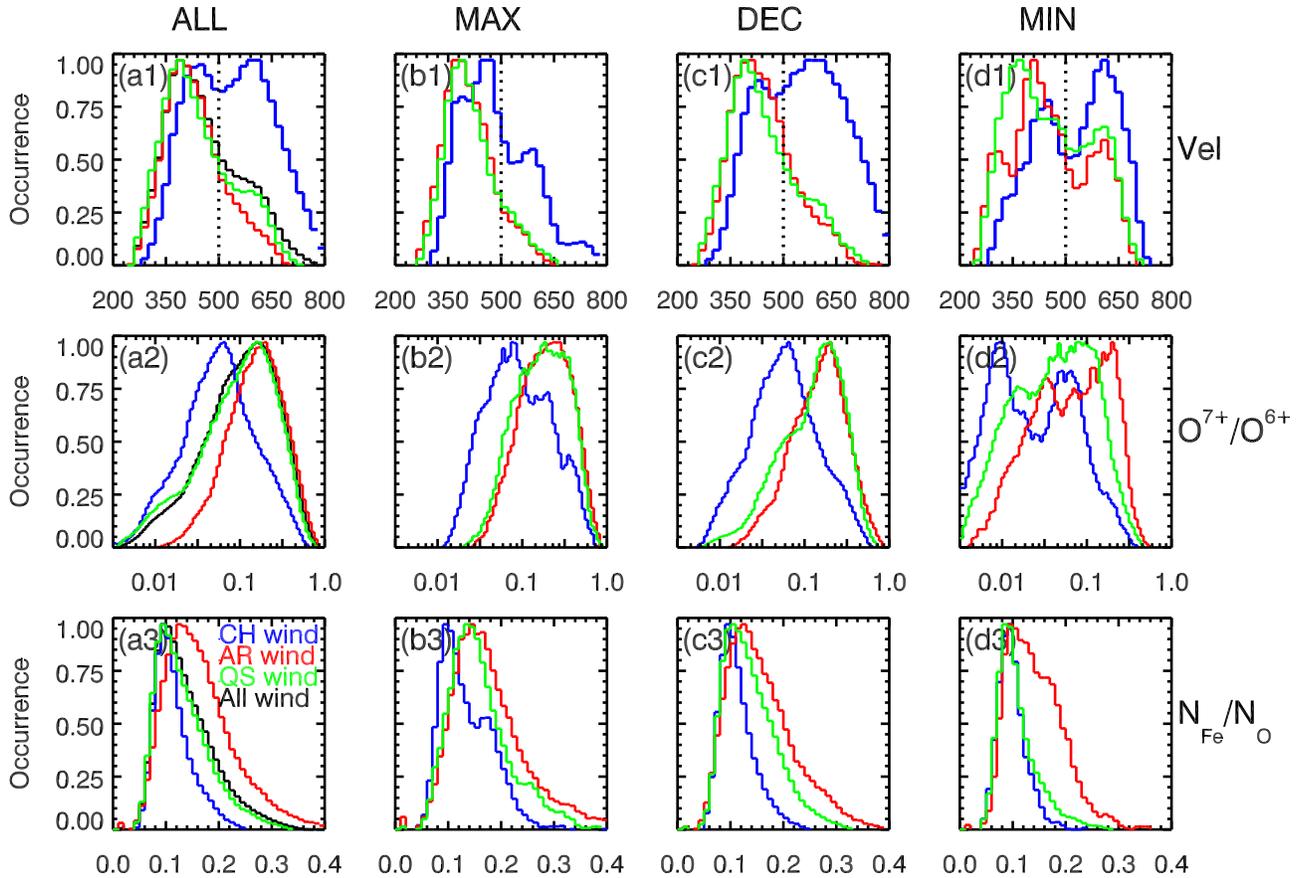}}
\caption{
The {\ {normalized}} distributions of the solar wind parameters.
The rows represent the parameters speed, \oql, and \feo\,
    while the first column is the solar wind as a whole,
    and the second (third, and fourth)
    column corresponds to the solar maximum (declining, and
    minimum) phase.
Blue, red, and green represents the CH, AR, and QS wind separately.
Although there are some differences in parameter distributions
    for different types of solar wind,
    it has a big over-lap.
The significant characteristic is a bimodal speed distribution
    for CH wind in all three solar phases.
\label{fig:fig_parameter_histrogm}}
\end{figure*}

\subsection{Distributions in the space of speed vs \oql}
\label{subsec:spacedistribution_vel_o76}

Figure~\ref{fig:fig_vel_o76_space} shows the {  scatter} plot of \oql\
    versus the solar wind speed for the different source regions,
    i.e. CH, AR, and QS.
For the purpose of being comparable to previous studies,
    we show in column 1 of Figure~\ref{fig:fig_vel_o76_space}
    the relation between \oql\ and the solar wind speeds
    for all three regions summed together.
The distributions for each individual source regions
    are shown in column 2, 3 and 4 of Figure~\ref{fig:fig_vel_o76_space}.
The anti-correlation of the solar wind speed and \oql\
    remains valid for all three types of solar wind.
In the left column of Figure~\ref{fig:fig_vel_o76_space},
    the linear fit of the distributions for the different source regions of
    solar wind are overplotted,
    showing that slopes and intercepts are the same
    for the different solar-wind source regions.
{\  Quantitatively, the slops are $-0.0027, -0.0028$ and $-0.0028$ for
    CH, AR and QS wind, respectively, during solar maximum,
    and $-0.0026, -0.0025, -0.0030$ at the decline phase,
    and $-0.0038, -0.0035, -0.0034$ during the solar minimum.
The absolute values of the slops for a certain type of
    solar wind are almost constant during the solar maximum and decline phases
    and larger during the solar minimum.
}

The anti-correlation between solar wind speed and \oql\ was first reported
    by observations from the \textit{International Sun-Earth Explorer 3} (ISEE~3)
    and Ulysses \citep{1989SoPh..124..167O, 1995SSRv...72...49G}.
\citet{2003JGRA..108.1158G} suggested that this observational fact
    supports the notion that the mechanism
    suggested by the RLO models is the main physical process  for heating and acceleration of the nascent solar wind.
Our statistical study shows that the anti-correlation
    between solar wind speeds and \oql\ is not only
    present for the CH wind
    \citep{2014ApJ...787..121K}
    but is also valid for the  AR and QS wind.
This suggests that the mechanisms which account for the anti-correlation
    between the solar wind speeds and \oql\ ratio are the same
    in CH, QS, and AR wind.
Relating to the RLO models, this means that the correlation between
    loop size and loop temperature
    is similar in all {  three types of regions}.
With regard to the WTD models, this suggests that the super radial expansion and
    curvature degree of the open magnetic filed lines
    is proportional to the source region temperature.
{\  The anti-correlation between the solar wind speed and \oql\
    can also be explained by
    a scaling law, in which a  higher coronal electron temperature
    leads to more energy lost from radiation for SSW,
    and vice versa as suggested by
    \citet{2003ApJ...599.1395S}, \citet{2014JGRA..119.1486S}, and \citet{2011ApJ...739....9S}.
 This scaling law is required for all
    magnetically driven solar wind models.
Considering that the physical parameters and magnetic field configurations
    for a certain type of  region (CH, AR and QS)
    have a large range,
    our statistical results for the different types of  solar wind provide a test
    for this scaling law.
Our results demonstrate that the relationship for solar wind speed and \oql\
    is valid and is almost the same
    for different types of solar wind,
    which means the scaling law is valid in all three types of solar wind.
}

\begin{figure*}
\figurenum{3}
\centerline{\includegraphics[width=0.95\textwidth,clip=]
    {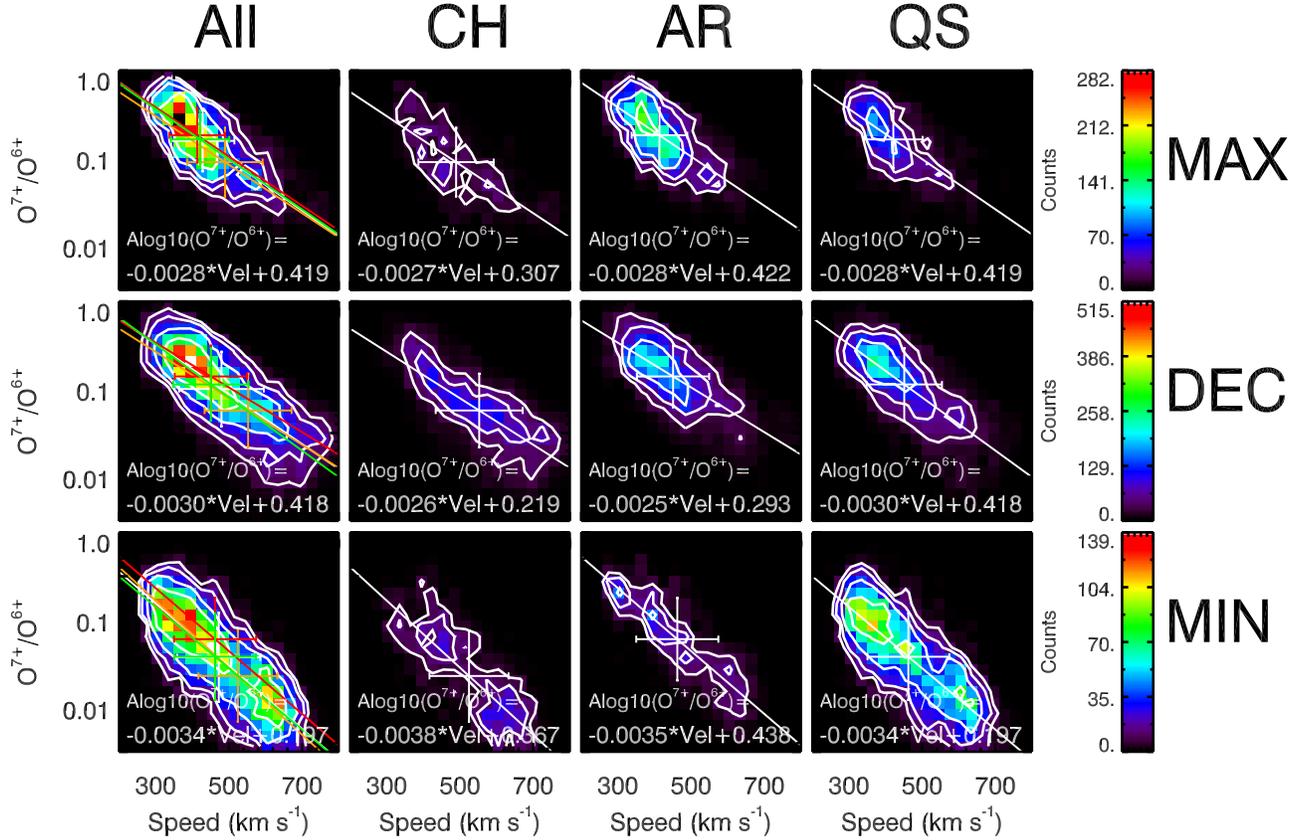}}

\caption{
The distribution of solar wind originating in CH, AR, and QS regions
    in the speed--\oql\ space.
The first to fourth column represents the wind as a whole,
    and from CH, AR, and QS regions, respectively,
    while the top (middle, bottom) panel corresponds to the solar maximum
    (declining, minimum) phase.
In the left column, the white line represents the linear fitting result
    for the solar wind as a whole,
    and orange (red, green) represents the wind coming from CH (AR, QS) regions.
The slopes and intercepts are given in the bottom of each figure.
The error bars represent the standard deviations of speed
    and \oql ratio.
The counts mean the number of hourly solar wind units
    detected by the ACE satellite.
\label{fig:fig_vel_o76_space}}
\end{figure*}

\subsection{Distributions in the space of speed vs \feo\ }
\label{subsec:spacedistribution_vel_feo}
Figure~\ref{fig:fig_vel_FeO_space} presents the {  scatter} plot of the
    solar wind speed versus  \feo\,
    again for all three regions together in column~1,
    and for the individual source regions in columns~2, 3, and 4.
The distribution for all three regions together is similar to
    earlier studies
    \citep[e.g.][and the references therein]{2016SSRv..201...55A}.
There are four important features concerning the relation between
    \feo\ and the solar wind speeds.
First, the average value of \feo\ is the highest in the AR wind, and the lowest
    for the CH wind.
Second, the \feo\ range (0.06--0.40, FIP bias range 1--7) for the AR wind is wider than
    for the CH wind (0.06--0.20, FIP bias range 1--3).
Third, similar to the wind as a whole,
    the \feo\ ranges and their average values all decrease
    with the increasing solar wind speed in the different types of solar wind.
Fourth, the minimum value of \feo\ is similar ($\sim$0.06, FIP bias~$\sim$1)
    for all source regions and it does not change with the speed of the solar wind.

The remote measurements in the solar corona given by
    \citet{2001ApJ...555..426W}, \citet{2011ApJ...727L..13B}, and
    \citet{2013ApJ...778...69B}
    show that the FIP bias is higher
    in AR regions (dominated by loops) than in CHs.
The FIP bias in CHs is between 1 and 1.5
    \citep{2005JGRA..110.7109F},
    while in ARs
    it can reach values larger than 4 in the case of older ARs
    \citep{2001ApJ...555..426W}.
The remote measurements of the solar corona
    also show that the variation of FIP bias in ARs is larger than in CHs \citep{1992ApJ...389..764M, 2001ApJ...555..426W}.
Our results demonstrate that the \feo\ ranges and their average value are higher in the AR wind.
This means that the plasma stored in closed loops can escape into
    interplanetary space,
    and that this mass supply scenario is consistent with the RLO models.

The differences in \feo\ ranges and their average values between different
    source regions of solar wind can also be explained
    qualitatively by the RLO models.
{\  {\citet{2005JGRA..110.7109F} reviewed the morphological features in the
    upper atmosphere.
They showed that the small loops (10--20 arcsecs) are cooler (30\,000~K to 0.7~MK)
    and have shorter lifetime (100 to 500~s)
    in QS and CH regions.
There are also larger loops (tens to hundreds of arcsecs) which have
    higher temperature (1.2--1.6 MK) and longer lifetime (1--2 days) in QS regions.
By reconstructing the magnetic field with the help
    of a potential magnetic field model,
    \citet{2004SoPh..225..227W} suggest that the loops in CHs are
    on average flatter and shorter than in the QS.
The range of loop sizes and temperatures are wider in AR regions
    including small (10--20 arcsecs) cool ($<$0.1~MK) loops
    \citep{2015ApJ...810...46H}, as well  cool (0.1~MK -- 1~MK),
    warm (1--2M~K) and hot
    hot ($>$ 2 MK) loops with lengths ranging from
    a few tens to a few hundreds of arcsecs
    \citep[e.g.][Xie et al. ApJ (submitted)]{2004ApJ...608.1133L, 2008ApJ...680.1477A}.
}}
{  The above results mean that the ranges of temperatures and
    loop sizes are wider in AR regions than in CHs.
Although this relation is not strictly proportional,
    the lifetime of loops is connected to these parameters.}
\citet{2001ApJ...555..426W}
    studied the FIP bias of four emerging active regions
    and found that the FIP bias increases progressively after the emergence.
They concluded that the low FIP elements enrichment relates to
    the age of coronal loops.
The AR wind may both come from new loops (with low FIP bias)
    and old loops (with higher FIP bias).
Therefore, the \feo\ range and its average value in the AR wind is
    wider (higher) than in the CH wind.

The fact that \feo\ range and average value decrease with the
    increasing solar wind speed
    in all three types of solar wind
    (see Figure~\ref{fig:fig_vel_FeO_space})
    can also be explained qualitatively by the RLO models.
\citet{2003JGRA..108.1157F}
    showed that the speed of the solar wind is inverse
    to the loop {  temperatures}.
For the fast wind, the loops in the source regions are {  cooler} and their
    life time is shorter.
Thus, their FIP bias is lower and its distribution range is narrower.
In contrast, the slow wind is at the other extreme.
There are two possible interpretations for the fact that
    the minimum value of \feo\ is similar ($\sim$0.06, FIP bias~$\sim$1)
    for all source regions
    and it does not change with speed of the solar wind.
First, in the RLO models,
    some of the new born loops (size may be large or small)
    reconnect with the open field lines
    producing solar wind with lower \feo\ (wind speed is slow or fast).
{\  Second, based on the WTD models the solar wind escapes  directly
    along the open magnetic field lines
    and the FIP fractionation is restricted to
    the top of the chromosphere
    based on a model in which the FIP fractionation is caused
    by ponderomotive force
    \citep{2004ApJ...614.1063L, 2012ApJ...744..115L}.
Thus, the FIP bias is lower {($\sim$1--2)} in open filed lines
    compared with that in the closed loops ($\sim$2--7, please see Table 3 and 4 in \citet{2015LRSP...12....2L}).
In all {  three types of regions}, the solar wind which escapes from
    open magnetic filed
    lines directly has lower FIP bias, regardless of whether the speed is slow or fast.
}

\begin{figure*}
\figurenum{4}
\centerline{\includegraphics[width=0.95\textwidth,clip=]
    {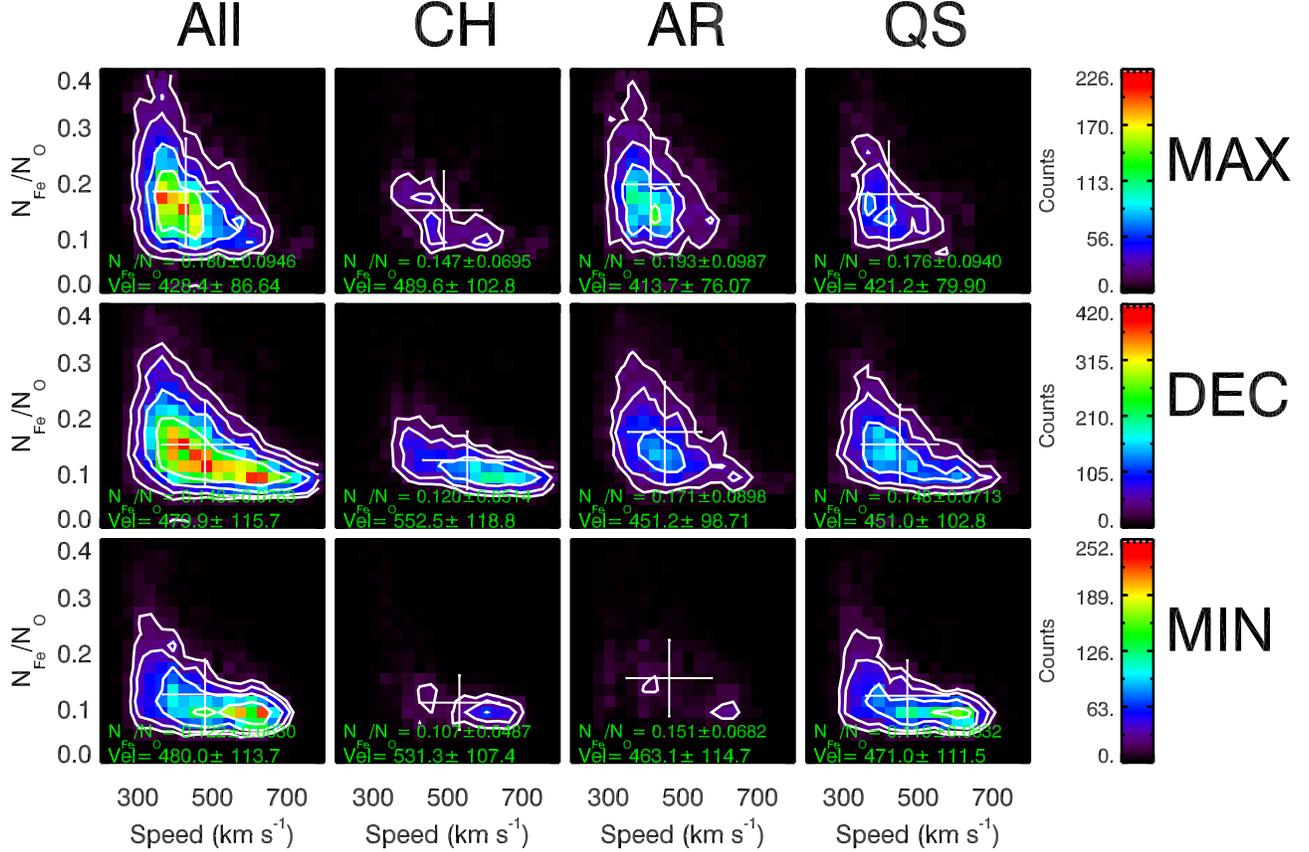}}
\caption{
Similar to Figure~\ref{fig:fig_vel_o76_space}, but for the distribution of
    solar wind in the speed--\feo\ space.
The overall behavior is similar for different source regions of solar wind.
The \feo\ range in AR wind is wider than in CH wind.
The minimum of \feo\ does not change with speed, and it is similar
    for different source regions of solar wind.
\label{fig:fig_vel_FeO_space}}
\end{figure*}

\section{Summary and concluding remarks}
\label{sec:conclusion}

The main purpose of this work was to examine
    the statistical properties of the solar wind
    originating from different solar regions, i.e. CHs, ARs, and QS.
The solar wind speeds, \oql, and \feo\
    were analyzed for different solar cycle phases
    (maximum, decline, and minimum).
Our main results can be summarized as follows:

\begin{enumerate}
\item
We found in the present study that
    the proportions of FSW and SSW are {  59.3\% and 40.7\%} for CH regions.
Fast solar wind is also found to emanate from AR and the QS,
    and the proportion of the FSW from AR and QS  with respect to their total solar wind input are
    13.7\% and 17.0\%, 25.8\% and 28.4\%,  34.0\% and 36.8\%,
    during the solar maximum, decline, and minimum phases, respectively.
The distributions of speed, \oql, and \feo\ ratio
    for the different source regions of solar wind have large overlaps
    indicating that it is hard to distinguish the source regions
    only by those wind parameters.
\item
We found that the speed distribution of the CH wind
    is bimodal in all three solar activity phases.
The peak of the fast wind from CHs for the period of time studied here is found to be at
    $\sim$~600 \velunit\ and {  the} slow wind peak is at $\sim$~400 \velunit.
The fast and slow wind components possibly come from
    the center and boundary regions of CHs, respectively.
\item
This study demonstrates that the anti-correlation between the speed and \oql\
    ratio remains valid in all three types of solar wind
    and during the three studied solar cycle activity phases.
\item
{  We identify four features of the distribution of
    \feo\ in the different solar wind types.
The average value of \feo\ is  highest in the AR wind,
    and  lowest for the CH wind.
The average values and ranges of \feo\ all
    decrease with the solar wind speed.
The \feo\ range in the AR wind is
    larger (0.06--0.40) than in CH wind (0.06--0.20).
The minimum value of \feo\ ($\sim$~0.06) does not change with
    the variation of speed,
    and it is similar for all source regions.}
\end{enumerate}

The statistical results indicate that
    the solar wind streams that come from different source regions
    are subject to similar constraints.
This suggests that the heating and acceleration mechanisms
    of the nascent solar wind in coronal holes, active regions,
    and quiet Sun have great similarities.
The two-peak distribution of the CH wind and
    the anti-correlation between the speed and \oql\
    in all three types of solar wind can be explained qualitatively
    by both the WTD and RLO models,
    whereas the distribution features of \feo\ in different source regions
    of solar wind can be explained more reasonably by the RLO models.

\acknowledgements{
{  The authors thank very much the anonymous referee for the helpful comments and suggestions.}
We thank the ACE SWICS, SWEPAM, and MAG instrument
    teams and the ACE Science Center for providing the ACE data.
SoHO is a project of international cooperation
    between ESA and NASA.
This research is supported by
    the National Natural Science Foundation of China (41274178,41604147,
    41404135, 41474150,
    41274176, and 41474149).
H.F. thanks the Shandong provincial Natural Science Foundation (ZR2016DQ10).
Z.H. thanks the Shandong provincial Natural Science Foundation (ZR2014DQ006)
   and the China Postdoctoral Science Foundation for financial supports.
}

\bibliographystyle{aasjournal}
\bibliography{classify}

\end{document}